\documentclass[aps, prd, twocolumn, nofootinbib]{revtex4-2}

\usepackage{amsmath}
\usepackage{amsfonts}
\usepackage{wasysym}
\usepackage{graphicx}
\usepackage{comment}
\usepackage{tensor}
\usepackage{hyphenat}

\def\filetype{pdf}

\def\path{}

\allowdisplaybreaks[1]

\usepackage{etoolbox}

\begin{document}



\title{Boson stars and their radial oscillations}
\author{Ben Kain}
\affiliation{Department of Physics, College of the Holy Cross, Worcester, Massachusetts 01610, USA}

\begin{abstract}
\noindent We review the derivation of the pulsations equations for spherically symmetric boson stars and then make a thorough study of the radial oscillation frequencies for the fundamental and first excited modes.  We do this for self\hyp interacting boson stars and consider a range of values for the self\hyp coupling constant.  We also numerically evolve boson stars and Fourier transform the dynamic solutions.  The Fourier transform gives an independent computation of the radial oscillation frequencies and allows us to verify our results obtained from the pulsation equations.  We find excellent agreement between the two methods.
\end{abstract}

\maketitle


\section{Introduction}

Boson stars are starlike configurations of a complex scalar field minimally coupled to gravity \cite{Kaup:1968zz, Ruffini:1969qy}.  They are described by classical solutions to the Einstein\hyp Klein\hyp Gordon system.  In their simplest realization, spacetime is spherically symmetric and time independent.  Such static solutions can be both stable and unstable to small perturbations.  To study the linear stability of boson stars, Gleiser \cite{Gleiser:1988rq}, Jetzer \cite{Jetzer:1988vr}, and Gleiser and Watkins \cite{Gleiser:1988ih} derived pulsation equations, whose solution gives the squared radial oscillation frequency, which can be used to determine stability (see also \cite{Lee:1988av}).  Alternatively, both linear and nonlinear stability can be studied by numerically evolving a boson star using full numerical relativity, as was first done by Seidel and Suen \cite{Seidel:1990jh} (see also \cite{Balakrishna:1997ej, Hawley:2000dt}).  
For reviews on boson stars, see \cite{Jetzer:1991jr, Schunck:2003kk, Liebling:2012fv}.

Stability has proven to be a powerful motivation for the development of tools and methods for studying boson stars.  These tools and methods, however, can do more than just tell us about stability.  Our interest in this work is to make a thorough study of the radial oscillation frequencies of boson stars.  We shall do this for boson stars that include self-interactions for the scalar field.

The study of radial oscillations of compact objects was initiated by Chandrasekhar when he derived a pulsation equation for spherically symmetric systems with a perfect fluid energy-momentum tensor \cite{Chandrasekhar:1964zz}.  His pulsation equation has since been used extensively to study the radial oscillations of neutrons stars (see, for example, \cite{Chanmugam, Gondek:1997fd, Kokkotas:2000up, Brillante:2014lwa, Sagun:2020qvc}).  One motivation for studying radial oscillations is to gain insight into the inner structure of the star.

We have another motivation, which is to help with the study of dark matter admixed neutron stars \cite{Henriques:1989ar, Sandin:2008db, Leung:2011zz}.  It is possible that dark matter could be mixed with the ordinary nuclear matter inside a neutron star.  If sufficient amounts of dark matter exist inside the star, bulk properties such as mass, radius, and radial oscillation frequencies are affected \cite{Comer:1999rs, Leung:2011zz, Leung:2012vea, ValdezAlvarado:2012xc, Kain:2020zjs, Kain:2021hpk}.  Dark matter admixed neutrons stars with bosonic dark matter (also known as fermion-boson stars) have been studied semianalytically \cite{Henriques:1989ar, Henriques:1989ez, Henriques:1990xg} and with full numerical relativity \cite{ValdezAlvarado:2012xc, Brito:2015yfh, Valdez-Alvarado:2020vqa, DiGiovanni:2020frc}. To understand the radial oscillations of the mixed systems, it is useful to have an understanding of the radial oscillations of the individual systems.  Hence, there is a need to study the radial oscillations of boson stars.

The pulsation equations for boson stars were first solved numerically by Gleiser and Watkins \cite{Gleiser:1988ih}.  Their interest was in determining the onset of instability and they did not solve for the radial oscillation frequencies.  This was extended by Hawley and Choptuik \cite{Hawley:2000dt}, who did compute the radial oscillation frequencies for the fundamental and first excited modes, though their interest was primarily with black hole critical phenomena and they did not include self-interactions for the scalar field.  Recently, radial oscillation frequencies were computed for $\ell$-boson stars in \cite{Alcubierre:2021mvs}.  Nonradial oscillations of boson stars were considered in \cite{Kojima:1991np, Yoshida:1994xi, Macedo:2013jja}.  Radial oscillations for a pseudo-Goldstone boson was studied in \cite{Lopes:2019eue}.

In this work, we both reproduce and extend the results of \cite{Hawley:2000dt}.  We compute the radial oscillation frequencies for boson stars for a range of values for the self-coupling constant.  As will be seen, the pulsation equations are complicated.  It is valuable, then, to have an independent method of computing the radial oscillation frequencies so as to verify that the pulsation equations have been derived and solved correctly.  We therefore numerically evolve boson stars using full numerical relativity.  Fourier transforming the dynamic solutions gives an independent computation of the radial oscillation frequencies.  We find excellent agreement between the two methods.

In the next section, we review the spherically symmetric Einstein-Klein-Gordon system and give the general set of equations that describe boson stars.  In the sections that follow, we rewrite these equations in various ways.  In Sec.\ \ref{sec:static}, we reduce them to describing static, or equilibrium, solutions, in which spacetime is time independent.  We then solve the static equations for the well\hyp known static boson star solutions.  In Sec.\ \ref{sec:radial}, we write the equations to first order in perturbations to the static solutions.  We then form the pulsation equations, which we solve for the squared radial oscillation frequencies.  Lastly, in Sec.\ \ref{sec:dynamic}, we solve the equations using full numerical relativity.  We then Fourier transform the result and compare with the radial oscillation frequencies computed in Sec.\ \ref{sec:radial}.  We conclude in Sec.\ \ref{sec:conclusion}.  In Appendix \ref{sec:freq}, we list some radial oscillation frequencies and in Appendix \ref{sec:comparsion}, we show that our results are consistent with those in \cite{Gleiser:1988ih, Hawley:2000dt}. In Appendix \ref{app:units}, we list some convenient unit conversions.


\section{Spherically symmetric Einstein-Klein-Gordon system}
\label{sec:EKG}

In this paper, we restrict our attention to spherically symmetric systems and use units such that $c=\hbar = 1$.  We parametrize the spherically symmetric metric as
\begin{equation} \label{metric}
ds^2 = -e^{\nu} dt^2 + e^{\lambda} dr^2 +r^2 d\theta^2 + r^2 \sin^2\theta \, d\phi^2,
\end{equation}
where $\nu(t,r)$ and $\lambda(t,r)$ are determined from the Einstein field equations,
\begin{equation} \label{GR}
G\indices{^\mu_\nu} = 8\pi G T\indices{^\mu_\nu},
\end{equation}
where $G\indices{^\mu_\nu}$ is the Einstein tensor, which is determined from the metric in Eq.\ (\ref{metric}), $T\indices{^\mu_\nu}$ is the energy-momentum tensor, which is given below, and $G = 1/\sqrt{m_P}$, where $m_P$ is the Planck mass.  In the following sections, we shall make use of the $(\mu,\nu) = (t,t), \, (r,r)$, and $(t,r)$ equations from (\ref{GR}), which lead to 
\begin{equation} \label{metric eqs}
\begin{split}
\nu' &= +8\pi G r e^{\lambda} T\indices{^r_r} + \frac{e^{\lambda}-1}{r}
\\
\lambda' &= 
-8\pi G r e^\lambda T\indices{^t_t}
- \frac{e^\lambda - 1}{r} 
\\
\dot{\lambda} &= -8\pi G r e^\nu T\indices{^t_r},
\end{split}
\end{equation}
where a dot denotes a $t$ derivative and a prime denotes an $r$ derivative.

For the matter sector, we use the standard Lagrangian for a complex scalar field,
\begin{equation} \label{L}
\mathcal{L} = - (\partial_\mu \phi) (\partial^\mu \phi)^* - \mu^2 |\phi|^2 - \eta |\phi|^4,
\end{equation}
where $\phi(t,r)$ is the complex scalar field, $\mu$ is its mass, and $\eta \geq 0$ is the self-coupling constant.  We minimally couple the scalar field to gravity through $\mathcal{L} \rightarrow \sqrt{-g} \mathcal{L}$, where $g$ is the determinant of the metric.  From this Lagrangian it is straightforward to compute the equations of motion,
\begin{equation} \label{eom}
\begin{split}
\partial_t \left[ e^{(\lambda-\nu)/2} \dot{\phi} \right] &= \frac{1}{r^2} \partial_r \left[ r^2 e^{(\nu-\lambda)/2} \phi' \right]
\\
&\qquad
- e^{(\nu+\lambda)/2} \left( \mu^2 + 2 \eta |\phi|^2 \right) \phi,
\end{split}
\end{equation}
and the following components of the energy\hyp momentum tensor,
\begin{equation} \label{em comps}
\begin{split}
T\indices{^t_t} &= - e^{-\nu} |\dot{\phi}|^2 - e^{-\lambda} |\phi'|^2 - \mu^2|\phi|^2 - \eta |\phi|^4
\\
T\indices{^r_r} &= + e^{-\nu} |\dot{\phi}|^2 + e^{-\lambda} |\phi'|^2 - \mu^2|\phi|^2 - \eta |\phi|^4
\\
T\indices{^t_r} &= - e^{-\nu} \left( \dot{\phi} \phi^{*\prime} + \dot{\phi}^* \phi' \right).
\\
\end{split}
\end{equation}
There are also nonzero values for $T\indices{^\theta_\theta} = T\indices{^\phi_\phi}$, which we will not be using, and $T\indices{^r_t} = -e^{\nu - \lambda} T\indices{^t_r}$. 

The Lagrangian in (\ref{L}) is invariant under global phase transformations, $\phi \rightarrow e^{i\theta} \phi$, for constant $\theta$.  This leads to a conserved current,
\begin{equation}
J^\mu = i g^{\mu\nu} (\phi^*\partial_\nu \phi - \phi \partial_\nu \phi^*),
\end{equation}
which satisfies a continuity equation, $\nabla_\mu J^\mu = 0$.  This continuity equation immediately leads to a conserved charge.  From $q(t,r) = - \int_0^r d^3r \sqrt{-g} J^t$, we have
\begin{equation} \label{Q}
q' = -4\pi r^2 e^{(\nu + \lambda)/2} J^t.
\end{equation}
The solution to Eq.\ (\ref{Q}) gives $q(t,r)$ and the conserved charge is given by $Q = q(t,r\rightarrow \infty)$.

This section has presented the general set of equations that we will use to study boson stars.  In the following sections, we will solve these equations for static solutions, for radial oscillation frequencies, and for dynamic solutions.


\section{Static solutions}
\label{sec:static}

In this section, we reduce the equations given in Sec.\ \ref{sec:EKG} to those that describe static, or equilibrium, solutions.  Static solutions are solutions for which the geometry is time independent.  For the geometry to be time independent, the energy\hyp momentum tensor must be time independent and diagonal.  It is not difficult to show that this requirement on the energy\hyp momentum tensor requires the scalar field to take the form
\begin{equation} \label{ansatz}
\phi(t,r) =\phi_0(r) e^{-i\omega t},
\end{equation}
where $\omega$ is a real constant and $\phi_0$ is real up to a global phase.  Since the Lagrangian in (\ref{L}) is invariant under global phase transformations, without loss of generality we take $\phi_0$ to be real.  Equation (\ref{ansatz}) is often referred to as the \textit{boson star ansatz}.  We shall indicate time independent fields with a subscript 0.

Assuming (\ref{ansatz}) and dropping the time dependence of the metric fields, the equations of motion in (\ref{eom}) reduce to 
\begin{equation} \label{static eom}
\begin{split}
0 &= \phi_0'' + \left( \frac{2}{r} + \frac{\nu'_0}{2} - \frac{\lambda'_0}{2} \right) \phi_0'
\\
&\qquad
+ e^{\lambda_0} \left( \omega^2 e^{-\nu_0} - \mu^2 - 2 \eta \phi_0^2 \right) \phi_0,
\end{split}
\end{equation}
the energy\hyp momentum tensor components in (\ref{em comps}) reduce to
\begin{equation} \label{equil em comps}
\begin{split}
T^t_{0t}
&= 
- \omega^2 e^{-\nu_0} \phi_0^2 
- e^{-\lambda_0} \phi_0^{\prime \, 2}
- \mu^2 \phi_0^2 - \eta \phi_0^4 
\\
T^r_{0r}
&= 
+ \omega^2 e^{-\nu_0} \phi_0^2
+ e^{-\lambda_0} \phi_0^{\prime \, 2}
- \mu^2 \phi_0^2 - \eta \phi_0^4,
\end{split}
\end{equation}
along with $T^t_{0r}= 0$, and the metric equations in (\ref{metric eqs}) become
\begin{equation} \label{static metric eqs}
\begin{split}
\nu_0' &= +8\pi G r e^{\lambda_0} T^r_{0r} + \frac{e^{\lambda_0}-1}{r}
\\
\lambda_0' &= 
-8\pi G r e^{\lambda_0} T^t_{0t}
- \frac{e^{\lambda_0} - 1}{r},
\end{split}
\end{equation}
with the bottom equation in (\ref{metric eqs}) vanishing identically.  A useful equation that we will make use of in the following section is
\begin{equation} \label{useful metric eq}
\lambda_0' + \nu_0' = 
16\pi G r \left(
\phi_0^{\prime \, 2}
+ \omega^2 \phi_0^2 e^{\lambda_0-\nu_0} 
\right),
\end{equation}
where we plugged in for the energy\hyp momentum tensor components using (\ref{equil em comps}).  Finally, the equation for the conserved charge in (\ref{Q}) becomes 
\begin{equation} \label{equil Q}
q_0' = 8\pi \omega r^2 \phi_0^2 e^{(\lambda_0 - \nu_0)/2}.
\end{equation}

Static boson stars are described by the solutions to Eqs.\ (\ref{static eom}) and (\ref{static metric eqs}), with the conserved charge given by the solution to Eq.\ (\ref{equil Q}). To solve these equations, it is convenient to define $\sigma_0(r)$ and $m_0(r)$ through
\begin{equation} \label{sigma m}
\sigma_0 \equiv e^{(\nu_0 + \lambda_0)/2}, \qquad
m_0 \equiv \frac{r}{2G} \left( 1- e^{-\lambda_0} \right).
\end{equation}
We further define $\Phi_0(r) \equiv \phi_0'(r)$ so that we have a system of first order ODEs,
\begin{equation} \label{static eqs}
\begin{split}
\Phi_0' &=  
-\left[ \frac{2}{r} + \frac{4\pi G r}{N_0} \left( T^r_{0r} + T^t_{0t} \right) + \frac{2 G m_0}{N_0 r^2} \right] \Phi_0
\\
&\qquad - \frac{1}{N_0} \left( \frac{\omega^2}{N_0\sigma^2} - \mu^2 - 2 \eta \phi_0^2 \right) \phi_0
\\
\sigma_0' &= 4\pi G \frac{ r \sigma_0}{N_0} \left( T^r_{0r} - T^t_{0t} \right)
\\
m_0' &= -4\pi r^2 T^t_{0t}
\\
q_0' &= 8\pi \frac{\omega r^2 \phi_0^2}{N_0 \sigma_0},
\end{split}
\end{equation}
along with $\phi_0' = \Phi_0$, where $N_0 \equiv 1 - 2 G m_0/r$ .  

These equations can be numerically integrated outward from $r=0$ after inner boundary conditions are determined.  Inner boundary conditions can be determined by plugging Taylor expansions of the fields into Eqs.\ (\ref{static eqs}), giving
\begin{align} \label{iBC}
\phi_0(r) &= \phi_0(0) - \frac{\phi_0(0)}{6} \left[ \frac{\omega^2}{\sigma_0^2(0)}  - \mu^2 - 2 \eta \phi_0^2(0) \right] r^2 + O(r^4)
\notag \\
\sigma_0(r) &= \sigma_0(0) + \frac{4 \pi G \omega^2 \phi_0^2 (0)}{\sigma_0(0)} r^2 + O(r^4),
\end{align}
along with $m_0(r) = O(r^3)$ and $q_0(r) = O(r^3)$.  As it stands,  $\sigma_0(0)$ and $\omega$ are unknown.  It greatly simplifies finding solutions to define
\begin{equation} \label{sigma omega hat}
\hat{\sigma}_0(r) \equiv \frac{\sigma_0(r)}{\sigma_0(0)}, \qquad
\hat{\omega} \equiv \frac{\omega}{\sigma_0(0)},
\end{equation}
and then to replace all occurrences of $\sigma_0$ and $\omega$ with $\hat{\sigma}_0$ and $\hat{\omega}$.  The advantage in doing this is that $\sigma_0(0)$ cancels out and the inner boundary condition for $\hat{\sigma}_0$,
\begin{equation} \label{sigma iBC}
\hat{\sigma}_0(r) = 1 + 4 \pi G \hat{\omega}^2 \phi_0^2 (0) r^2 + O(r^4),
\end{equation}
does not require knowledge of $\sigma_0(0)$.  Outer boundary conditions are obtained by requiring that the energy\hyp momentum tensor goes to zero at large $r$, and so $\phi_0, \Phi_0\rightarrow 0$ as $r \rightarrow \infty$.  We further require that the spacetime is asymptotically Schwarzschild, so that $\sigma_0 \rightarrow 1$, and thus that $\hat{\sigma}_0 \rightarrow 1 / \sigma_0(0)$, as $r\rightarrow \infty$.

In this paper, we consider only fundamental static solutions.  Such solutions have zero nodes (i.e\ zero\hyp crossings for $\phi_0$) and are uniquely identified by the central value $\phi_0(0)$ and the self-coupling $\eta$.  To find a solution, we choose values for $\phi_0(0)$ and $\eta$ as well as a trial value for $\hat{\omega}^2$ and then integrate Eqs.\  (\ref{static eqs}) outward from $r = 0$ using the inner boundary conditions in Eqs.\ (\ref{iBC}) and (\ref{sigma iBC}).  We then use the shooting method, varying the value of $\hat{\omega}^2$ until our integrated solution satisfies the outer boundary conditions.  Once a solution is found, the mass and conserved charge of the system are given by $M = m_0(\infty)$ and $Q = q_0(\infty)$ and the squared oscillation frequency is given by $\omega^2 = \hat{\omega}^2/\hat{\sigma}^2(\infty)$.  Solutions are only found for positive $\omega^2$ and hence only for real $\omega$.

\begin{figure*}
\centering
\includegraphics[width=6.75in]{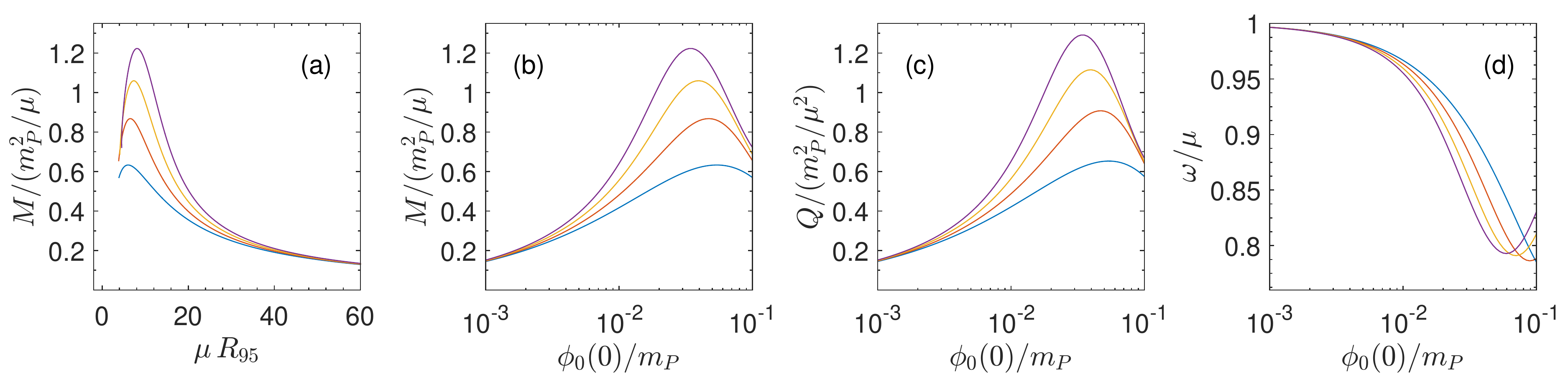}
\caption{The well\hyp known static boson star solutions are presented for four values of the self-coupling constant $\eta$.  (a) The mass, $M$, of the boson star as a function of $R_{95}$, where $R_{95}$ is the radius that contains 95\% of the mass.  (b)--(c) The mass, conserved charge $Q$, and oscillation frequency $\omega$ as a function of the central value $\phi_0(0)$.  From bottom to top in (a)--(c) and top to bottom in (d), the curves are for $\eta / (\mu^2/m_P^2) = 0$ (blue), 100 (red), 200 (orange), and 300 (purple).}
\label{fig:static_MRomega}
\end{figure*}

In Fig.\ \ref{fig:static_MRomega}, we present solutions for a few values of the self-coupling $\eta$.  Figure \ref{fig:static_MRomega}(a) displays the mass, $M$, of the boson star as a function of its radius, where $R_{95}$ is defined as the radius that contains 95\% of the mass.  Figures \ref{fig:static_MRomega}(b)--\ref{fig:static_MRomega}(c) display the mass, conserved charge $Q$, and oscillation frequency $\omega$ as a function of the central value $\phi_0(0)$.

In this work, we display all results using dimensionless quantities that are standard for boson stars, such as $M/(m_P^2/\mu)$ and  $\mu R_{95}$ for the boson star's mass and radius (see, for example, \cite{Gleiser:1988rq, Jetzer:1988vr, Gleiser:1988ih, Seidel:1990jh}).  The benefit in doing this is that all results are valid for an arbitrary scalar field mass, $\mu$, and $\mu$ does not have to be specified.  In Appendix \ref{app:units}, we list a few astrophysical\hyp friendly unit conversions, which do require specification of the scalar field mass.

\begin{figure}
\centering
\includegraphics[width=2.85in]{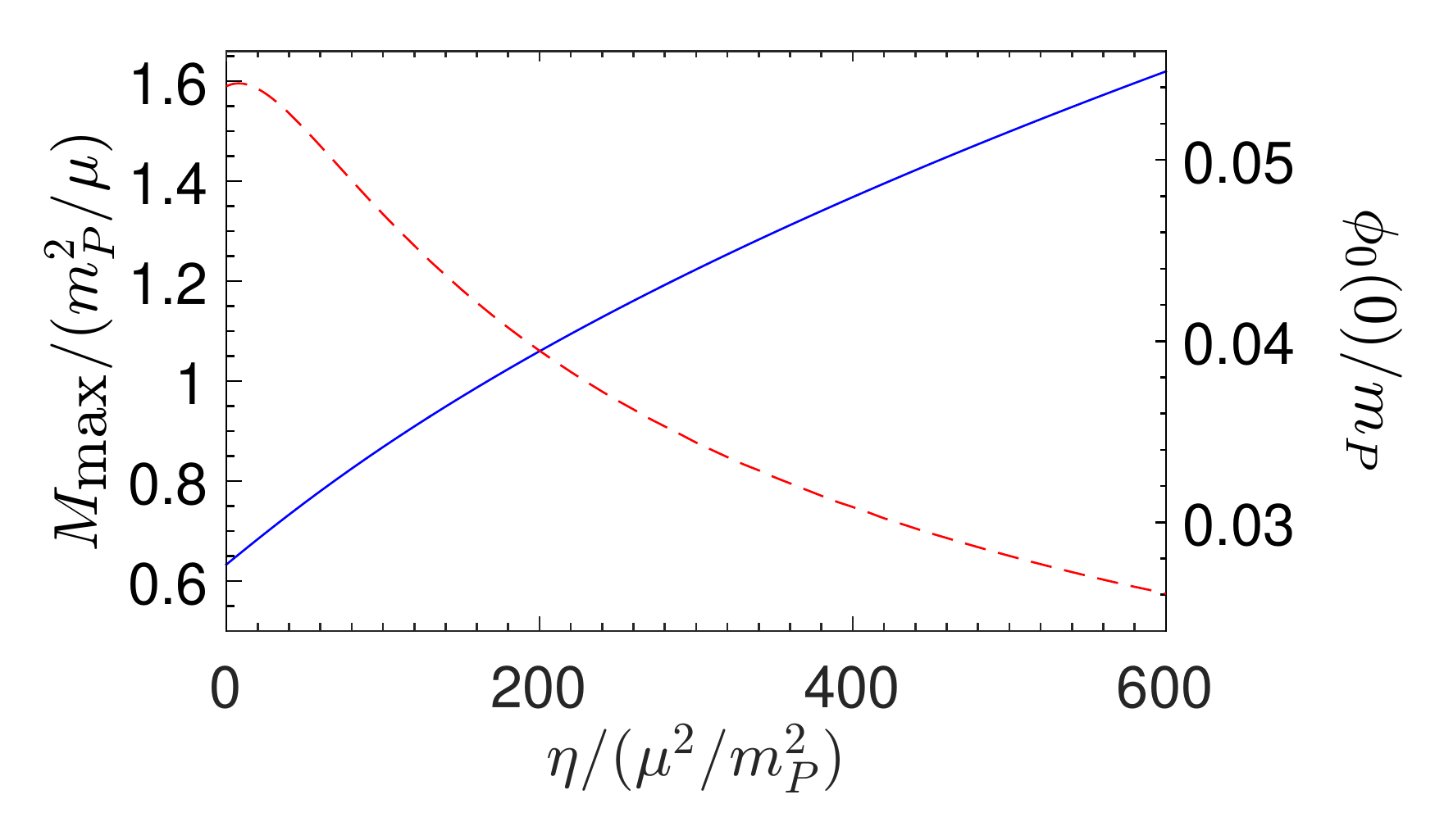}
\caption{The solid blue curve, which uses the vertical scale on the left, plots the maximum mass of a static solution for a given value of $\eta$.  The dashed red curve, which uses the vertical scale on the right, plots the central value $\phi_0(0)$ corresponding to the maximum mass.  Since the transition from stable to unstable for static solutions occurs at the static solution with the largest mass, the dashed red curve is also plotting the critical value of $\phi_0(0)$, above which a static solution is unstable.}
\label{fig:Mmax}
\end{figure}

From Fig.\ \ref{fig:static_MRomega}, we can see that for a given value of $\eta$, there is a maximum possible mass.  It is well\hyp known that the transition from stable to unstable with respect to small perturbations occurs at the static solution with the largest mass \cite{ShapiroBook, Liebling:2012fv}.  In Fig.\ \ref{fig:Mmax}, the solid blue curve, which uses the vertical scale on the left, gives the maximum mass and the dashed red curve, which uses the vertical scale on the right, plots the central value $\phi_0(0)$ corresponding to the maximum mass.  Both quantities are plotted as a function of $\eta$.  The dashed red curve, then, is also plotting the critical value of $\phi_0(0)$, above which the boson star is unstable.



\section{Radial oscillations}
\label{sec:radial}

Gleiser \cite{Gleiser:1988rq}, Jetzer \cite{Jetzer:1988vr}, and Gleiser and Watkins \cite{Gleiser:1988ih} were the first to derive pulsation equations for boson stars.  These papers studied the same system, but Ref.\ \cite{Gleiser:1988ih} combined the equations differently than Refs.\ \cite{Gleiser:1988rq, Jetzer:1988vr} and, in this sense, derived different pulsation equations.  We prefer the formalism of \cite{Gleiser:1988rq, Jetzer:1988vr}, as it is more easily able to accommodate all fields, and we review that here.  We find it easiest to compare equations with \cite{Gleiser:1988ih} and we therefore define the perturbations as done in \cite{Gleiser:1988ih}.  Our numerical method for solving the pulsation equations generalizes the method presented in  \cite{Gleiser:1988ih}. 


\subsection{Pulsation equations}
\label{sec:pulsation}

In this subsection, we write the equations in Sec.\ \ref{sec:EKG} to first order in perturbations about the static solutions and derive the pulsation equations for boson stars.  Since a complex scalar field has two real degrees of freedom, we should expect two coupled pulsation equations.  The solutions to the pulsation equations give the squared radial oscillation frequencies.

We begin by defining 
\begin{equation}
\phi(t,r) \equiv \left[\phi_1(t,r) + i\phi_2(t,r) \right]e^{-i\omega t},
\end{equation}
where $\omega$ is the oscillation frequency for a static solution and $\phi_1$ and $\phi_2$ are real.  Note that setting $\phi_2 = 0$ (and dropping the time dependence) gives the boson star ansatz (\ref{ansatz}).  Consequently, the static value of $\phi_2$ vanishes and $\phi_2$ is at the level of a perturbation.  We next write the fields as perturbations about their static solutions, defining the perturbations as in \cite{Gleiser:1988ih},
\begin{equation} \label{perturbatons}
\begin{split}
\phi_1(t,r) &= \phi_0(r) \left[ 1 + \delta \phi_1(t,r) \right]
\\
\phi_2(t,r) &= \phi_0(r) \delta \phi_2(t,r)
\\
\nu(t,r) &= \nu_0(r) + \delta\nu(t,r)
\\
\lambda(t,r) &= \lambda_0(r) + \delta\lambda(t,r).
\end{split}
\end{equation}

Writing the fields in the equations of motion in (\ref{eom}) as perturbations about their static solutions, then keeping perturbations only through first order and canceling the static terms, we find
\begin{equation} \label{phi1 eom}
\begin{split}
0 &= \delta \phi_1''
+ \delta \phi_1' \left( 2 \frac{\phi_0'}{\phi_0} + \frac{2}{r} + \frac{\nu_0' - \lambda_0'}{2} \right)
+ \frac{\delta\nu' - \delta\lambda'}{2} \frac{\phi_0'}{\phi_0}
\\
&\qquad
- e^{\lambda_0 - \nu_0} (2\omega \delta \dot{\phi}_2 + \delta \ddot{\phi}_1 )
 +e^{\lambda_0 - \nu_0} \omega^2 (\delta \lambda -  \delta \nu)
\\
&\qquad
- e^{\lambda_0} \mu^2 \delta \lambda
- 2 \eta \phi_0^2 e^{\lambda_0} \left( 2  \delta \phi_1 +  \delta \lambda \right),
\end{split}
\end{equation}
and 
\begin{equation} \label{phi2 eom}
\begin{split}
0 &= \delta\phi_2''
+\left(2 \frac{\phi_0'}{\phi_0} + \frac{2}{r} + \frac{\nu_0' - \lambda_0'}{2} \right) \delta\phi_2'
-  \omega e^{\lambda_0 - \nu_0} \frac{\delta \dot{\nu} - \delta\dot{\lambda}}{2}
\\
&\qquad
- e^{\lambda_0 - \nu_0} \left( \delta \ddot{\phi}_2 - 2 \omega \delta \dot{\phi}_1 \right)
- 2 \eta e^{\lambda_0} \phi_0^2 \delta \phi_2,
\end{split}
\end{equation}
where we used Eq.\ (\ref{static eom}) to cancel terms.  Doing the same for the energy\hyp momentum tensor components in (\ref{em comps}) gives
\begin{equation}\label{delta Ttt}
\begin{split}
\delta T\indices{^t_t}
&= \omega^2 e^{-\nu_0} \phi_0^2  \delta \nu
+ e^{-\lambda_0} \phi_0^{\prime 2} \delta \lambda
- 2\omega^2 e^{-\nu_0} \phi_0^2  \delta \phi_1
\\
&\qquad
+ 2\omega e^{-\nu_0} \phi_0^2 \delta \dot{\phi}_2 
- 2 e^{-\lambda_0} \phi_0^{\prime \, 2} \delta \phi_1
\\
&\qquad
- 2 e^{-\lambda_0} \phi_0 \phi_0' \delta \phi_1'
- 2\mu^2 \phi_0^2 \delta \phi_1
 - 4\eta \phi_0^4 \delta \phi_1
\\
\delta T\indices{^r_r} &= - \delta T \indices{^t_t} - 4 \mu^2 \phi_0^2 \delta \phi_1 -8 \eta \phi_0^4 \delta \phi_1
\\
\delta T\indices{^t_r}
&= -2 e^{-\nu_0} \phi_0  \left( \phi_0' \delta \dot{\phi}_1 - \omega \phi_0 \delta \phi_2' \right)
\end{split}
\end{equation}
and for the metric equations in (\ref{metric eqs}) gives
\begin{equation} \label{delta metric eqs}
\begin{split}
\delta \nu' 
&= +8\pi G r e ^{\lambda_0} \delta T\indices{^r_r}  + \nu_0' \delta \lambda + \frac{\delta \lambda}{r}
\\
\delta \lambda'
&= -8\pi G r e^{\lambda_0}\delta T\indices{^t_t} 
+ \lambda_0' \delta \lambda - \frac{\delta \lambda}{r}
\\
\delta\dot{\lambda} &= -8\pi G r e^{\nu_0} \delta T\indices{^t_r},
\end{split}
\end{equation}
where the first two metric equations were simplified by using the static metric equations in (\ref{static metric eqs}).  Two useful formulas that follow from Eqs.\ (\ref{delta metric eqs}) are
\begin{equation} \label{useful delta eqs}
\begin{split}
\partial_r\left(r e^{-\lambda_0} \delta \lambda \right) &= -8\pi G r^2 \delta T\indices{^t_t}
\\
\delta \nu' - \delta \lambda'
&= \left(\nu_0' - \lambda_0' + \frac{2}{r} \right) \delta \lambda
\\
&\qquad
-32 \pi G r e^{\lambda_0}\phi_0^2 ( \mu^2   +2 \eta \phi_0^2  )\delta \phi_1,
\end{split}
\end{equation}
where in the bottom equation we plugged in for the perturbed energy-momentum tensor using Eqs.\ (\ref{delta Ttt}).  Finally, for the conserved charge in (\ref{Q}),
\begin{equation} \label{delta Q}
\delta q' = 8\pi \omega r^2 \phi_0^2 e^{(\lambda_0 - \nu_0)/2} \left( 2 \delta \phi_1 + \frac{\delta \lambda - \delta \nu}{2} - \frac{\delta \dot{\phi}_2}{\omega}  \right).
\end{equation}

There are different ways to combine the equations presented thus far.  Further, not all of the equations are necessary, since they are not all independent.  In deriving the pulsation equations, we follow the method of \cite{Gleiser:1988rq, Jetzer:1988vr}.  We introduce the quantity
\begin{equation} \label{xi def}
\dot{\xi} \equiv \omega \, \delta \phi_2,
\end{equation}
where the factor of $\omega$ is included for convenience.  Upon plugging this into the bottom equation in (\ref{delta metric eqs}), we obtain a result that can be immediately integrated to give
\begin{equation} \label{delta lambda}
\delta \lambda = 16 \pi G r \phi_0 
\left( \phi_0' \delta \phi_1 - \phi_0 \xi' \right).
\end{equation}
Combining this result, $\delta T\indices{^t_t}$ in Eq.\ (\ref{delta Ttt}), and the equilibrium equations (\ref{static eom}) and (\ref{useful metric eq}) with the top equation in (\ref{useful delta eqs}) allows us to solve for $\delta \nu$,
\begin{align} \label{delta nu}
\delta \nu &= 
- \frac{2}{\omega^2}  \ddot{\xi}
+ \frac{2}{\omega^2} e^{\nu_0-\lambda_0} \xi''
+ \left(
 16\pi G  r \phi_0 \phi_0' 
+
4 
\right) \delta\phi_1 
\notag \\
&\qquad
+ \frac{2}{\omega^2}  e^{\nu_0-\lambda_0}  
\left(\frac{2}{r} + 2\frac{\phi_0'}{\phi_0} - \lambda_0' 
+ 8\pi G r \phi_0^{\prime \, 2}
\right)
\xi'.
\end{align}

We now define
\begin{equation}
\zeta(t,r) \equiv \xi'(t,r).
\end{equation}
We also introduce a harmonic time dependence for all perturbations:
\begin{equation}
\delta \phi_1(t,r) = \delta \phi_1(r) e^{-i\chi t}, \qquad
\zeta(t,r) = \zeta(r) e^{-i\chi t}, \qquad
\end{equation}
$\delta \nu(t,r) = \delta \nu(r) e^{-i\chi t}$, and $\delta \lambda(t,r) = \delta\lambda(r) e^{-i\chi t}$, where $\chi$ is the radial oscillation frequency we would like to solve for.  The pulsation equations will be two coupled ODEs for the perturbations $\delta \phi_1$ and $\zeta$.  To construct the $\delta \phi_1$ pulsation equation, we combine the equation of motion in (\ref{phi1 eom}) with the bottom equation in (\ref{useful delta eqs}) and with Eqs.\ (\ref{delta lambda}) and (\ref{delta nu}) to obtain 
\begin{widetext}
\begin{equation}  \label{pulsation delta phi1}
\begin{split}
\delta \phi_1'' &= 
- \chi^2 e^{\lambda_0 - \nu_0} \delta \phi_1 
- \left( 2 \frac{\phi_0'}{\phi_0} + \frac{2}{r} + \frac{\nu_0' - \lambda_0'}{2} \right) \delta \phi_1' 
+2  \left(\frac{2}{r} + 2\frac{\phi_0'}{\phi_0} - \lambda_0' 
+ 8\pi G r \phi_0^{\prime \, 2}
\right) \zeta
+ 2\zeta'
\\
&\qquad
- 16 \pi G r \phi_0 
\left( \phi_0' \delta \phi_1 - \phi_0 \zeta \right)
\left[
\frac{1}{2} \frac{\phi_0'}{\phi_0} \left(\nu_0' - \lambda_0' + \frac{2}{r} \right)
+ e^{\lambda_0} \left(\omega^2 e^{- \nu_0} - \mu^2 - 2\eta \phi_0^2 
\right)
\right]
\\
&\qquad
+  \left[
4 \omega^2 e^{\lambda_0 - \nu_0}
+ 16 \pi G r e^{\lambda_0}\phi_0 \phi_0' ( 
\omega^2 e^{-\nu_0}
+\mu^2   +2 \eta \phi_0^2  )
 + 4\eta \phi_0^2 e^{\lambda_0}
\right] \delta \phi_1.
\end{split}
\end{equation}
For the $\zeta$ ODE, we begin with the bottom equation in (\ref{useful delta eqs}) and plug into it Eq.\ (\ref{delta lambda}), its derivative, and the derivative of Eq.\ (\ref{delta nu}) to obtain
\begin{align} \label{pulsation zeta}
\zeta'' &=
- \chi^2 e^{\lambda_0- \nu_0} \zeta
- 2 \omega^2 e^{\lambda_0 - \nu_0} 
 \delta\phi_1' 
-
\left(\frac{2}{r} + 2\frac{\phi_0'}{\phi_0} - \lambda_0' 
+ 8\pi G r \phi_0^{\prime \, 2}
\right) 
\left[ (\nu_0' - \lambda_0') \zeta
+ \zeta'
\right]
\notag \\
&\qquad 
- 
\left[ 2 \frac{\phi_0''}{\phi_0} - 2 \frac{\phi_0^{\prime\,2}}{\phi_0^2} - \frac{2}{r^2} - \lambda_0'' + 8\pi G \left( \phi_0^{\prime\,2} + 2r\phi_0' \phi_0'' \right) \right] \zeta
-(\nu_0' - \lambda_0') \zeta'
\\
&\qquad
-
8\pi G \omega^2 e^{\lambda_0-\nu_0}
\left[
 \left( \phi_0^2 + 2r \phi_0 \phi_0' \right) \zeta 
+  r \phi_0^2 \zeta'
-
r \phi_0 \left(\nu_0' - \lambda_0' + \frac{2}{r} \right)
( \phi_0' \delta \phi_1 - \phi_0 \zeta)
+ 
2  r e^{\lambda_0} \phi_0^2 ( \mu^2   +2 \eta \phi_0^2  )\delta \phi_1 \right].
\notag 
\end{align}
\end{widetext}
It is straightforward to show that Eqs.\ (\ref{pulsation delta phi1}) and (\ref{pulsation zeta}) are equivalent to equations (41) and (42) in \cite{Gleiser:1988rq} by using Eqs.\ (\ref{static eom}), (\ref{useful metric eq}), and the derivative of Eq.\ (\ref{useful metric eq}).

It turns out that $\delta q'$ in (\ref{delta Q}) can be integrated exactly.  Using Eqs.\ (\ref{xi def}), (\ref{delta lambda}), (\ref{delta nu}), and (\ref{useful metric eq}) in Eq.\ (\ref{delta Q}), we obtain \cite{Gleiser:1988rq, Jetzer:1988vr}
\begin{equation} \label{delta Q exact}
\delta q = - \frac{8\pi}{\omega} r^2 \phi^2 e^{(\nu_0 - \lambda_0)/2} \zeta.
\end{equation}

In Eqs.\ (\ref{pulsation delta phi1}) and (\ref{pulsation zeta}), $\nu_0'$ and $\lambda_0'$ are given by Eqs.\ (\ref{static metric eqs}), $\phi_0''$ is given by Eq.\ (\ref{static eom}), $\phi_0$, $\phi_0'$, $\nu_0$, $\lambda_0$, and $\omega$ are obtained from the static solution, and $\lambda_0''$ is obtained by taking the derivative of the $\lambda_0'$ equation in (\ref{static metric eqs}):
\begin{equation}
\lambda_0'' =
\frac{ e^{\lambda_0} - 1}{r^2}
- \lambda_0' \frac{e^{\lambda_0}}{r}
-8\pi G e^{\lambda_0}\left(
 T^t_{0t} +
   r \lambda_0'  T^t_{0t}
   +
   r \partial_r T^t_{0t} \right),
\end{equation}
where
\begin{equation}
\begin{split}
\partial_r T^t_{0t} 
&=  2 e^{-\lambda_0} \phi_0^{\prime \, 2} 
\left( \frac{2}{r} + \frac{\nu'_0}{2} \right)
- 4 \phi_0  \phi_0'\left(  \mu^2 + 2\eta \phi_0^2 \right) 
\\
&\qquad
+ \omega^2 \nu_0' e^{-\nu_0} \phi_0^2 .
\end{split}
\end{equation}

Solving the pulsation equations (\ref{pulsation delta phi1}) and (\ref{pulsation zeta}) gives $\delta \phi_1$ and  $\zeta$.  We can use these, along with the static solutions $\phi_0$, $\phi_0'$, $\nu_0$, and $\lambda_0$ to construct the perturbation to $\phi_1$, which from (\ref{perturbatons}) is $\phi_0 \delta\phi_1$, to obtain $\delta \lambda$ from (\ref{delta lambda}), which after writing it in terms of $\zeta$ is
\begin{equation} \label{delta lambda 2}
\delta \lambda = 16 \pi G r \phi_0 
\left( \phi_0' \delta \phi_1 - \phi_0 \zeta \right),
\end{equation}
and to obtain $\delta q$ from (\ref{delta Q exact}).  In the next subsection, $\phi_0 \delta \phi_1$, $\delta \lambda$, and $\delta q$ will play a role in how the pulsation equations are solved.


\subsection{Numerical solution and results}
\label{sec:results}

In this subsection, we outline how we numerically solve the pulsation equations for the squared radial oscillation frequencies, $\chi^2$, and present results.  Our method generalizes the strategy of Gleiser and Watkins \cite{Gleiser:1988ih} and is similar to that used by Hawley and Choptuik \cite{Hawley:2000dt}.  The relevant equations are the pulsation equations in (\ref{pulsation delta phi1}) and (\ref{pulsation zeta}), the $\delta q$ equation in (\ref{delta Q exact}), the $\delta \lambda$ equation in (\ref{delta lambda 2}), and the static equations in (\ref{static eqs}).  Since both (\ref{delta Q exact}) and (\ref{delta lambda 2}) are algebraic, once the other equations are solved, it is trivial to obtain $\delta q$ and $\delta \lambda$.

To solve the pulsation equations, we need inner and outer boundary conditions.  Note that all relevant equations that contain perturbations, contain one perturbed field in each of their terms.  Thus, we scale these equations such that $\delta \phi_1(0) = 1$.  Additional inner boundary conditions are found by plugging Taylor expansions of the fields into the pulsation equations, giving
\begin{align}
\delta \phi_1(r) &= 1 + \frac{1}{6} \left[6\zeta_1 + \frac{4\omega^2 - \chi^2}{\sigma^2_0(0)} +4 \eta \phi_0^2(0) \right]r^2 + O(r^4)
\notag \\
\zeta(r) &= \zeta_1 r + O(r^3),
\end{align}
where $\zeta_1$ is an as-yet-undetermined constant.

For outer boundary conditions we have that perturbations head to zero at large $r$.  In particular we shall use that $\phi_0 \delta \phi_1, \delta q, \delta \lambda \rightarrow 0$ as $r\rightarrow \infty$, where $\phi_0 \delta \phi_1$ is the definition of the $\phi_1$ perturbation from (\ref{perturbatons}).  The condition on $\delta q$ can be understood as only allowing perturbations that conserve total charge \cite{Gleiser:1988rq, Jetzer:1988vr, Gleiser:1988ih, Hawley:2000dt}.

We next write all equations in terms of $m_0$, $\hat{\sigma}_0$, and $\hat{\omega}$, as defined in Eqs.\ (\ref{sigma m}) and  (\ref{sigma omega hat}), and
\begin{equation}
\hat{\chi} \equiv \frac{\chi}{\sigma_0(0)}.
\end{equation}
Just as with the static equations, $\sigma_0(0)$ cancels out.  We note that it is for this reason that we defined $\dot{\xi}$ in Eq.\ (\ref{xi def}) with the factor of $\omega$.  

We assume that the static equations have been solved and the value of $\omega$ determined for given values of $\phi_0(0)$ and $\eta$.  Even so, we find it easiest to simultaneously solve the static equations (again) and the pulsation equations, but now with the precise value of $\omega$ plugged in.  We do this by writing the pulsation equations in first order form and then integrating outward from $r = 0$ using the previously listed inner boundary conditions and using trial values for $\hat{\chi}^2$ and $\zeta_1$.  We then use the shooting method, varying both $\hat{\chi}^2$ and $\zeta_1$ until the outer boundary conditions are satisfied.  In practice, we find that if two of the outer boundary conditions are satisfied (i.e.\ two of $\phi_0 \delta\phi_1, \delta q, \delta \lambda \rightarrow 0$ as $r\rightarrow \infty$), so is the third.  Once a solution is found, the squared radial oscillation frequency is given by $\chi^2 = \hat{\chi}^2/\hat{\sigma}^2_0(\infty)$.

The pulsation equations in (\ref{pulsation delta phi1}) and (\ref{pulsation zeta}) can be shown to be self-adjoint \cite{Gleiser:1988rq}.  We expect the squared radial oscillation frequencies to be real and discrete and, for a given static solution, for there to be an infinite number of them, just as for fermion stars \cite{Kokkotas:2000up}.  The smallest frequency is called the fundamental mode, the next smallest the first excited mode, and so on.  Such modes are also indicated  by the number of nodes (i.e.\ zero crossings) of either $\delta q$ or $\delta \lambda$ \cite{Gleiser:1988ih}.  In practice, we solve for a specific mode by tuning $\hat{\chi}^2$ and $\zeta_1$ in the shooting method such that $\delta q$ and $\delta \lambda$ have the desired number of nodes.  We present results for the fundamental and first excited modes.  We have attempted to solve for higher modes, but have been unable to do so.  We find that our code is running up against machine precision (we use 64\hyp bit floating point numbers).  Extending our code beyond this limitation is beyond the scope of this work.

\begin{figure}
\centering
\includegraphics[width=3.4in]{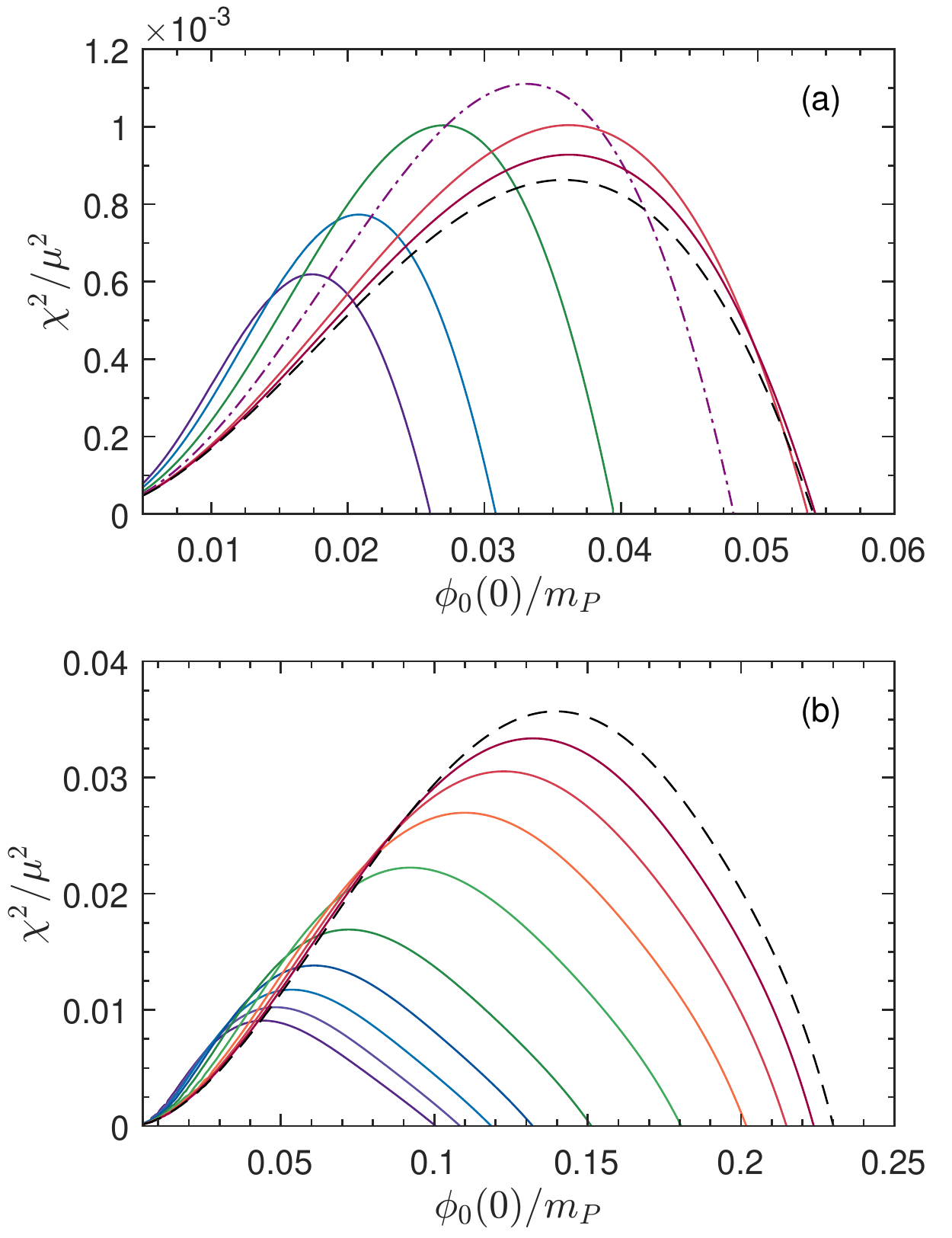}
\caption{The squared radial oscillation frequency, $\chi^2$, as a function of the central value $\phi_0(0)$ for (a) the fundamental mode and (b) the first excited mode.  The dashed curve in both plots corresponds to $\eta = 0$.  In (a), as $\eta$ is increased, the peak of each curve increases, until it hits a maximum at $\eta/(\mu^2/m_P^2) = 86.63$ (dash\hyp dotted curve).  As $\eta$ is increased further, the peaks decrease.  In (b), as $\eta$ is increased, the peaks always decrease.  In (a), the curves are for $\eta/(\mu^2/m_P^2) = 0$, 10, 25, 86.63, 200, 400, 600.  In (b) the curves are for $\eta/(\mu^2/m_P^2) = 0$, 10, 25, 50, 100, 200, 300, 400, 500, 600.}
\label{fig:radia_a}
\end{figure}

\begin{figure}
\centering
\includegraphics[width=3.5in]{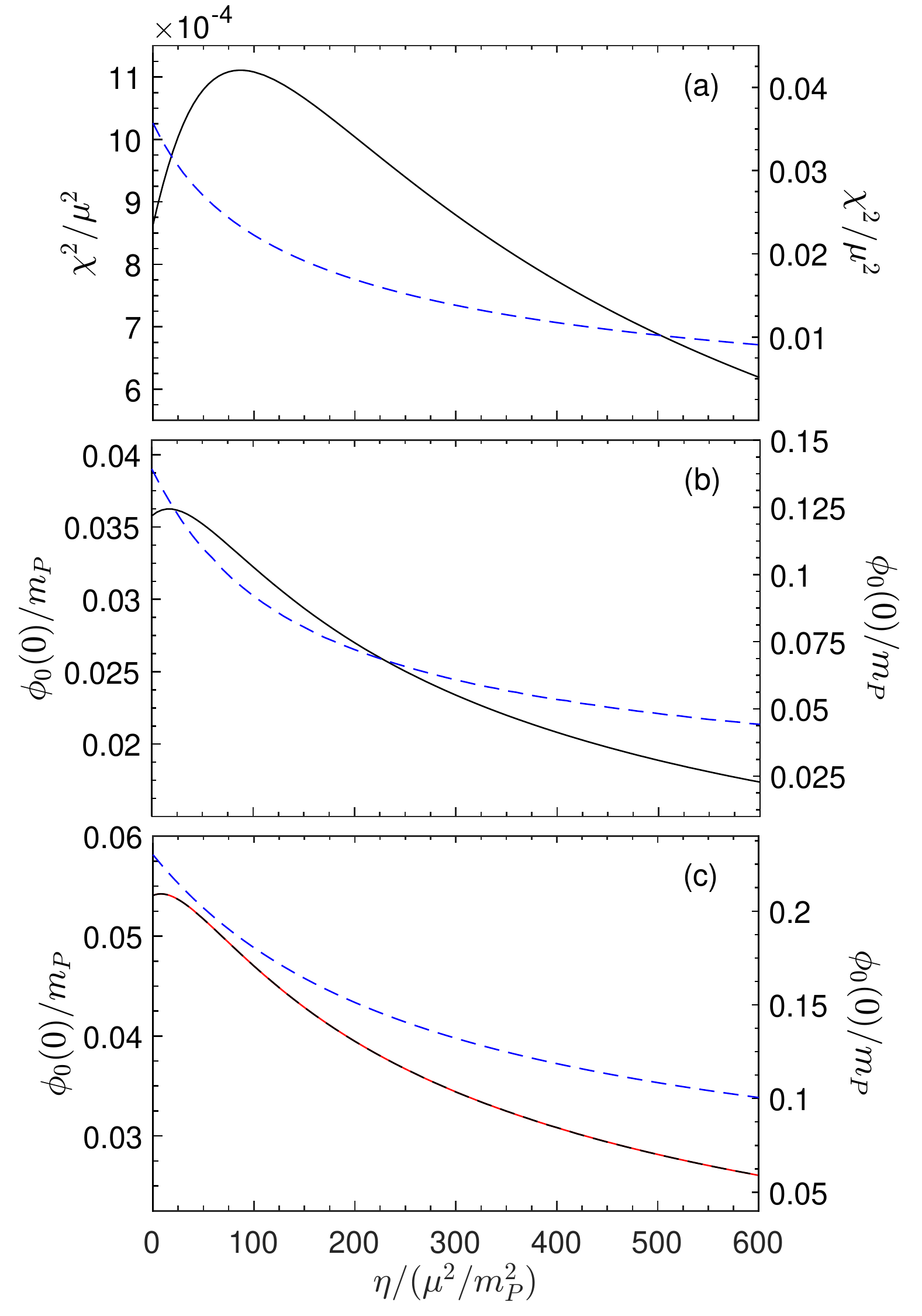}
\caption{In each plot, the solid black curve uses the vertical scale on the left and is for the fundamental mode and the dashed blue curve uses the vertical scale on the right and is for the first excited mode.  The maximum squared radial oscillation frequency for a given value of $\eta$ is plotted in (a) and the corresponding value of $\phi_0(0)$ is plotted in (b).  These values correspond to the peaks seen in Fig.\ \ref{fig:radia_a}.  (c) The central value $\phi_0(0)$ is plotted for when $\chi^2 = 0$.  This is the critical value of $\phi_0(0)$ and corresponds to when the respective mode transitions between stable and unstable.  The dashed red curve, which lies directly on top of the solid black curve, is the same dashed red curve plotted in Fig.\ \ref{fig:Mmax}.  That the solid black curve and the dashed red curve agree is expected, since both are computing the critical value of $\phi_0(0)$ for the fundamental mode. }
\label{fig:radia_b}
\end{figure}

In Fig.\ \ref{fig:radia_a}(a), we show the squared radial oscillation frequency, $\chi^2$, for the fundamental mode as a function of the central value $\phi_0(0)$ for various values of $\eta$.  As can be seen, for a given value of $\eta$, there exists a peak value of $\chi^2$.  As $\eta$ is increased from $\eta = 0$, the peak increases until it reaches a maximum value.  As $\eta$ is increased further, the peak decreases.  In Fig.\ \ref{fig:radia_a}(b), we show the analogous plot for the first excited mode.  We find again that for a given value of $\eta$, there exists a peak value of $\chi^2$.  Unlike with the fundamental mode, we find that as $\eta$ is increased, the peak always decreases.  We have computed the peak value of $\chi^2$ as a function of $\eta$ for both the fundamental and first excited modes, which is shown in Fig.\ \ref{fig:radia_b}(a).  In Fig.\ \ref{fig:radia_b}(b), we show the values of $\phi_0(0)$ at which the peak values of $\chi^2$ occur.

In Fig.\ \ref{fig:radia_a} we can see that all curves eventually hit zero for sufficiently large $\phi_0(0)$.  The transition from positive to negative for $\chi^2$ indicates the respective mode transitioning from stable to unstable.  If any mode is unstable, the static solution is unstable.  Since the fundamental mode always has the smallest frequency, a static solution is stable if $\chi^2$ for the fundamental mode is positive, otherwise the static solution is unstable.  The value of $\phi_0(0)$ when $\chi^2 = 0$, and hence when a mode is transitioning from stable to unstable, is called the critical value.  We have computed the critical value of $\phi_0(0)$ for both the fundamental and first excited modes.  We have done this by setting $\chi^2 = 0$ and varying $\zeta_1$ and $\phi_0(0)$ in the shooting method.  The results are shown in Fig.\ \ref{fig:radia_b}(c).  Also shown in Fig.\ \ref{fig:radia_b}(c) is the same dashed red curve as shown in Fig.\ \ref{fig:Mmax}, which gives the value of $\phi_0(0)$ for the static solution with the largest mass.  As mentioned in the previous section, it is well\hyp known that the static solution with the maximum mass occurs for the critical value of $\phi_0(0)$.  Consequently, the dashed red curve is directly on top of the solid black curve as expected.  Though expected, this is a nontrivial test of our radial oscillation code. 

In Appendix \ref{sec:freq}, we list radial oscillation frequencies and other computed values.  Some of the $\eta = 0$ results presented in this section were previously computed in \cite{Gleiser:1988ih, Hawley:2000dt}.  We show that our results are consistent with those in \cite{Gleiser:1988ih, Hawley:2000dt} in Appendix \ref{sec:comparsion}.


\section{Dynamic solutions}
\label{sec:dynamic}

It is valuable to compute the radial oscillation frequencies using a different method than used in the previous section so as to verify that the equations and solutions of the previous section are correct.  This is particularly true given the complexity of the pulsation equations in (\ref{pulsation delta phi1}) and (\ref{pulsation zeta}).  In this section, we numerically evolve a spherically symmetric boson star using full numerical relativity, with the static solutions of Sec.\ \ref{sec:static} used as initial data.  It is well\hyp known that the discretization error inherent in a numerical evolution acts as a perturbation \cite{Seidel:1990jh}.  By Fourier transforming the dynamic solution, we obtain an independent computation of the radial oscillation frequencies and are able to verify that that we have computed them correctly in the previous section. 

We begin by putting the equations in Sec.\ \ref{sec:EKG} into a form better suited for numerical evolution.  Toward this end, it is convenient to use the metric functions $\alpha(t,r)$ and $a(t,r)$, defined by
\begin{equation}
\alpha \equiv e^{\nu/2}, \qquad a \equiv e^{\lambda/2}.
\end{equation}
Defining
\begin{equation}
\Phi \equiv \phi', \qquad \Pi \equiv \frac{a}{\alpha} \dot{\phi},
\end{equation}
the equations of motion in (\ref{eom}) can be written in first order form as
\begin{equation}
\dot{\Pi} = \frac{1}{r^2} \partial_r \left( r^2 \frac{\alpha}{a} \Phi \right) - \alpha a (\mu^2 + 2 \eta |\phi|^2 ) \phi.
\end{equation}
Writing the fields in terms of their real and imaginary parts,
\begin{equation}
\phi = \phi_1 + i \phi_2, \qquad
\Phi = \Phi_1 + i \Phi_2, \qquad
\Pi = \Pi_1 + i\Pi_2,
\end{equation}
the equations of motion become 
\begin{equation} \label{evo1}
\begin{split}
\dot{\phi}_1 &= \frac{\alpha}{a} \Pi_1 
\\
\dot{\Phi}_1 &=
\partial_r
\left(\frac{\alpha}{a} \Pi_1\right) 
\\
\dot{\Pi}_1 &=
\frac{1}{r^2} \partial_r
\left( \frac{r^2 \alpha}{a}\Phi_1\right)
- \alpha a
[\mu^2 + 2 \eta (\phi_1^2 + \phi_2^2) ] \phi_1
\end{split}
\end{equation}
and
\begin{equation} \label{evo2}
\begin{split}
\dot{\phi}_2 &= \frac{\alpha}{a} \Pi_2 
\\
\dot{\Phi}_2 &=
\partial_r
\left(\frac{\alpha}{a} \Pi_2 \right) 
\\
\dot{\Pi}_2 &=
\frac{1}{r^2} \partial_r
\left( \frac{r^2 \alpha}{a}\Phi_2 \right)
- \alpha a
[\mu^2 + 2 \eta (\phi_1^2 + \phi_2^2) ] \phi_2.
\end{split}
\end{equation}
The metric equations in (\ref{metric eqs}), when written in terms of the fields $\alpha$ and $a$, become
\begin{equation} \label{a alpha eqs}
\begin{split}
\alpha' &= +4\pi G r  \alpha a^2 T\indices{^r_r} + \frac{\alpha(a^2-1)}{2r}
\\
a' &= 
-4\pi G r a^3 T\indices{^t_t}
- \frac{a(a^2 - 1)}{2r} 
\\
\dot{a} &= -4\pi G r \alpha^2 a T\indices{^t_r}
\end{split}
\end{equation}
and the energy-momentum tensor components in (\ref{em comps}) become
\begin{equation} 
\begin{split}
T\indices{^t_t} 
&= -\frac{\Pi_1^2 + \Pi_2^2 + \Phi_1^2 + \Phi_2^2}{a^2}
- \mu^2(\phi_1^2 + \phi_2^2) - \eta(\phi_1^2 + \phi_2^2)^2
\\
T\indices{^r_r} 
&=+\frac{\Pi_1^2 + \Pi_2^2 + \Phi_1^2 + \Phi_2^2}{a^2}
- \mu^2(\phi_1^2 + \phi_2^2) - \eta(\phi_1^2 + \phi_2^2)^2
\\
T\indices{^t_r} 
&=
- \frac{2}{a \alpha} \left(\Pi_1 \Phi_1 + \Pi_2 \Phi_2 \right).
\end{split}
\end{equation}

For initial data, we use a static solution computed as described in Sec.\ \ref{sec:static}.  We assume the initial data is time symmetric (i.e.\ that it occurs at $t=0$), so that, given a static solution $\phi_0(r)$, $\Phi_0(r)$, $\sigma_0(r)$, $m_0(r)$, and $\omega$, we have
\begin{equation}
\phi_1(0,r) = \phi_0,
\qquad
\Phi_1(0,r) = \Phi_0, 
\qquad
\Pi_1(0,r) = 0,
\end{equation}
and 
\begin{equation}
\phi_2(0,r) = 0, 
\qquad
\Phi_2(0,r) = 0, 
\qquad
\Pi_2(0,r) 
= \frac{\omega}{N_0\sigma_0} \phi_0,
\end{equation}
where $N_0 = 1-2 G m_0/r$.

The boundary conditions for the metric fields are $a(t,0) = 1$ and $\alpha(t,r) = 1/a(t,r)$ for $r\rightarrow \infty$.  Since our computational domain does not extend to infinity, we allow matter fields to exit at the outer boundary using standard outgoing wave boundary conditions,
\begin{equation}
\begin{split}
\dot{\phi}_1 &= -\frac{\phi_1}{r} - \Phi_1, \quad
\dot{\Pi}_1 = - \frac{\Pi_1}{r} - \Pi_1', \quad
\Phi_1 = - \frac{\phi_1}{r} - \Pi_1,
\\
\dot{\phi}_2 &= -\frac{\phi_2}{r} - \Phi_2, \quad
\dot{\Pi}_2 = - \frac{\Pi_2}{r} - \Pi_2', \quad
\Phi_2 = - \frac{\phi_2}{r} - \Pi_1.
\end{split}
\end{equation}

Our code is second order accurate and evolves the evolution equations in (\ref{evo1}) and (\ref{evo2}) using the method of lines and third order Runge\hyp Kutta.  The spatial derivatives in the evolution equations are finite differenced using centered stencils.  We use the common practice of writing the spatial derivatives in the $\dot{\Pi}_1$ and $\dot{\Pi}_2$ evolution equations as $r^{-2}\partial_r = \partial_{r^3}$ before finite differencing \cite{AlcubierreBook}.  At each time step, we solve the first two equations in (\ref{a alpha eqs}) using second order Runge\hyp Kutta.  We use a uniform computational grid with $\Delta r = 0.005/\mu$ or $0.01/\mu$, $r_\text{max} = 100/\mu$, and $\Delta t / \Delta r = 0.5$ (we find that for larger values of $\phi_0(0)$ we need the more accurate $\Delta r = 0.005/\mu$ to obtain a precise match between the results of this and the previous sections).

We compute radial oscillation frequencies from dynamic solutions using a fast Fourier transform.  The results presented are from the Fourier transform of $|\phi| = \sqrt{\phi_1^2 + \phi_2^2}$ at the innermost grid point (either at $r = 0.0025/\mu$ or $0.005/\mu$).  We carry all evolutions out to $t = 2\times 10^4/\mu$, allowing for a large number of oscillations and an accurate determination of the radial oscillation frequencies.

As mentioned, the initial data is a static solution, which is uniquely identified by $\phi_0(0)$ and $\eta$.  Discretizaton error in the numerical evolution causes the static solution to shift slightly before quickly settling into a stable configuration.  As such, the dynamic evolution is for a static solution with a sightly different value of $\phi_0(0)$ (but the same value of $\eta$) than used to make the initial data.  For a proper matching of the radial oscillation frequencies as computed using the methods of this and the previous sections, a precise value for $\phi_0(0)$ must be determined.  We have found that simply averaging $|\phi|$ at the innermost grid point over the whole of the evolution gives a sufficiently accurate value.

\begin{figure}
\centering
\includegraphics[width=3in]{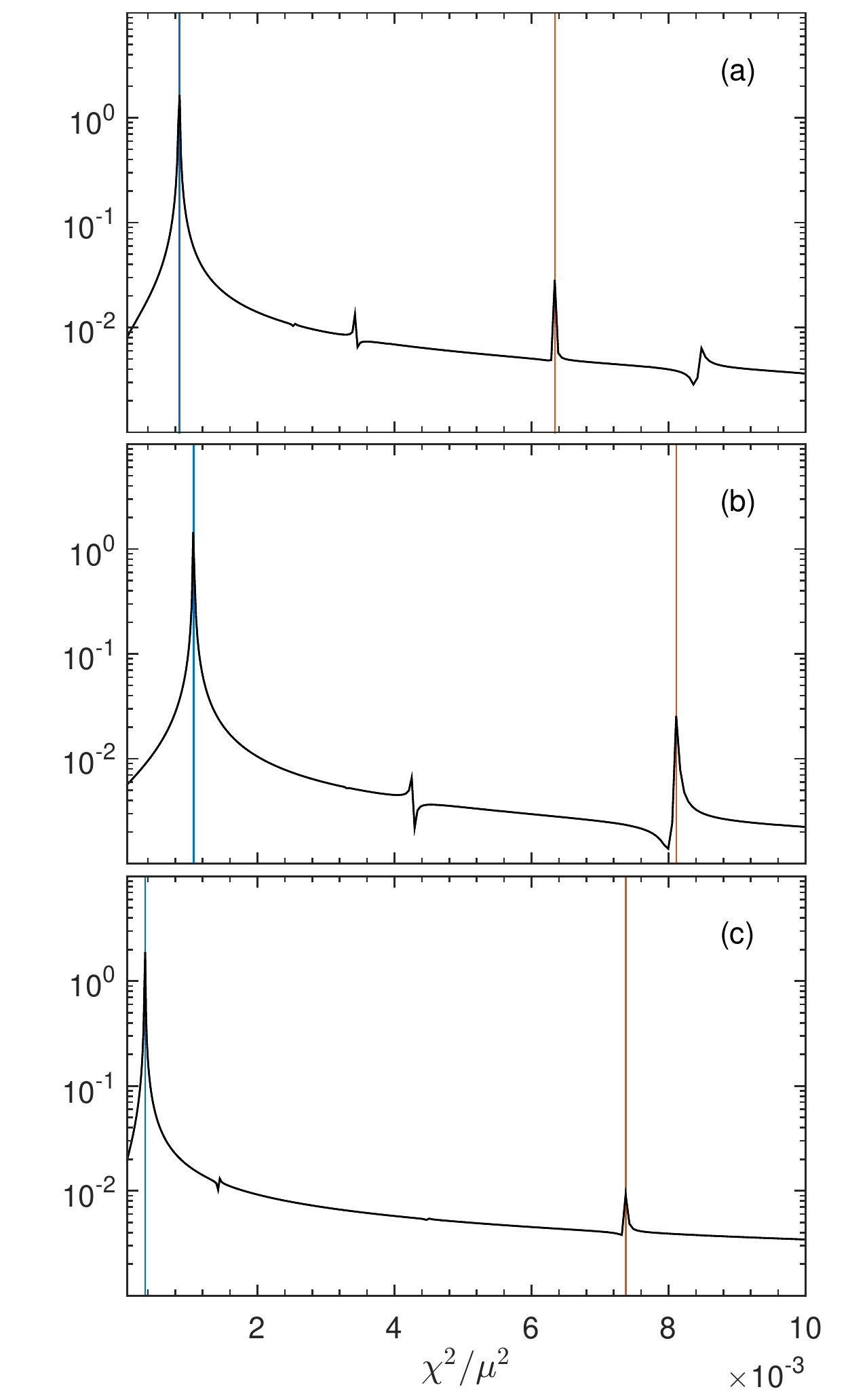}
\caption{Three representative plots are shown.  In each, the black curve is the Fourier transform of a dynamic solution and the spikes along the black curve give the frequencies at which the dynamic solution is oscillating.  The units on the vertical axis are arbitrary.  The vertical lines are the fundamental (left most, blue) and first excited (right most, red) squared radial oscillation frequencies computed from the Pulsation equations.  We can see that the agreement is excellent.  
(a) $\eta = 0$, $\phi_0(0)/m_P = 0.03522$, and radial oscillation frequencies $\chi^2/\mu^2 = 8.6222 \times 10^{-4}$ and $6.3409\times 10^{-3}$.  
(b) $\eta/(\mu^2/m_P^2) = 100$, $\phi_0(0)/m_P = 0.03517$, and radial oscillation frequencies $\chi^2/\mu^2 = 1.0729 \times 10^{-3}$ and $8.1150\times 10^{-3}$.
(c) $\eta/(\mu^2/m_P^2) = 400$, $\phi_0(0)/m_P = 0.02823$, and radial oscillation frequencies $\chi^2/\mu^2 = 3.6229 \times 10^{-4}$ and $7.3741\times 10^{-3}$.}
\label{fig:dynamic}
\end{figure}

In Fig.\ \ref{fig:dynamic}, we show three representative results.  The black curves are the Fourier transform of the dynamic solution and the spikes along the black curves represent frequencies at which the dynamic solution is oscillating.  It is not unreasonable to expect that there could be oscillation frequencies with a nonlinear origin.  As such, it is not necessarily clear which spikes have a linear origin and give the radial oscillation frequencies we are interested in.  The blue and red vertical lines are the fundamental and first excited squared radial oscillation frequencies as computed using the pulsation equations and the methods of the previous section.  As can can be seen, there is excellent agreement.  Thus, the solution to the pulsation equations allows us to unambiguously determine which of the spikes in Fig.\ \ref{fig:dynamic} come from linear perturbations, while the spikes in Fig.\ \ref{fig:dynamic} allow us to verify that we have computed the radial oscillation frequencies in the previous section correctly.

\section{Conclusion}
\label{sec:conclusion}

We computed the fundamental and first excited radial oscillation frequencies for self-interacting boson stars by solving pulsation equations.  To verify our results, we evolved boson stars using full numerical relativity and Fourier transformed the dynamic solutions.  We found excellent agreement between the two methods.  In Appendix \ref{sec:freq}, we list some radial oscillation frequencies and other computed values.

If boson stars are detectable, it is interesting to speculate on whether their radial oscillations can be measured.  Since radial oscillations do not couple to gravitational waves, it is unlikely that they can be measured by purely gravitational means.  For neutron stars, the expectation for how radial oscillations could be measured is through emission of electromagnetic radiation from the surface of the star (see, for example, \cite{Brillante:2014lwa, Sagun:2020qvc, Sun:2021cez}).  For something similar to be possible with boson stars, the scalar field would have to have additional interactions that were not considered in this work.

As mentioned in the Introduction, a main motivation for this work are dark matter admixed neutrons stars.  If sufficient amounts of bosonic dark matter mix with the ordinary nuclear matter of a neutron star, the radial oscillations of the mixed star can be affected \cite{Comer:1999rs, Leung:2011zz, Leung:2012vea, ValdezAlvarado:2012xc, Kain:2020zjs, Kain:2021hpk}.  This would occur for scalar field masses around $\mu  \sim 10^{-10}$ eV \cite{Henriques:1989ar}.  Boson stars made with such a scalar field have radial oscillation frequencies in the kHz range, which is comparable to neutron stars.  If the radial oscillations of mixed stars can be detected, say through emission of electromagnetic radiation, interesting questions are how the radial oscillations of the bosonic sector show up in the frequency spectrum \cite{ValdezAlvarado:2012xc} and if it is possible to infer details about the bosonic sector from their measurement.  If this latter case is possible, it would mean that the presence of dark matter could be inferred from measurements of the radial oscillation frequencies of neutron stars.  We expect the results presented here to be relevant for answering these questions, which are currently under study.


\appendix

\section{Radial oscillation frequencies}
\label{sec:freq}

In this appendix, we list some of the values computed in Sec.\ \ref{sec:radial}.  Table \ref{table freq node=0} lists values for the fundamental mode and Table \ref{table freq node=1} lists values for the first excited mode.

\setlength{\tabcolsep}{8pt}
\begin{table*}
\begin{tabular}{lllllll}
$\eta/(\mu^2/m_P^2)$ & $\phi_0(0)/m_P$ &$\omega/\mu$ & $\hat{\omega} / \mu$ & $\zeta_1/\mu^2$ &  $\chi^2/\mu^2$ & $\hat{\chi}^2/\mu^2$ \\
\hline
\hline
0&	 0.01&	 0.9668&	 1.0344&	 ${-0.7277}$&	 $0.1691\times 10^{-3}$&	 $0.1936\times 10^{-3}$\\
0&	 0.015&	 0.9513&	 1.0531&	 ${-0.7622}$&	 $0.3344\times 10^{-3}$&	 $0.4099\times 10^{-3}$\\
0&	 0.02&	 0.9365&	 1.0729&	 ${-0.7997}$&	 $0.5140\times 10^{-3}$&	 $0.6747\times 10^{-3}$\\
0&	 0.025&	 0.9224&	 1.0939&	 ${-0.8406}$&	 $0.6798\times 10^{-3}$&	 $0.9560\times 10^{-3}$\\
0&	 0.03&	 0.9089&	 1.1160&	 ${-0.8852}$&	 $0.8045\times 10^{-3}$&	 $1.2129\times 10^{-3}$\\
0&	 0.035&	 0.8961&	 1.1395&	 ${-0.9339}$&	 $0.8617\times 10^{-3}$&	 $1.3934\times 10^{-3}$\\
0&	 0.04&	 0.8840&	 1.1645&	 ${-0.9872}$&	 $0.8253\times 10^{-3}$&	 $1.4323\times 10^{-3}$\\
0&	 0.045&	 0.8724&	 1.1909&	 ${-1.0456}$&	 $0.6697\times 10^{-3}$&	 $1.2481\times 10^{-3}$\\
0&	 0.05&	 0.8615&	 1.2190&	 ${-1.1098}$&	 $0.3693\times 10^{-3}$&	 $0.7394\times 10^{-3}$\\
0&	0.05407&	0.8530&	1.2432&		${-1.1667}$&	0&	0\\
\hline
100&	 0.01&	 0.9635&	 1.0393&	 ${-0.7383}$&	 $0.2064\times 10^{-3}$&	 $0.2402\times 10^{-3}$\\
100&	 0.015&	 0.9442&	 1.0645&	 ${-0.7875}$&	 $0.4366\times 10^{-3}$&	 $0.5549\times 10^{-3}$\\
100&	 0.02&	 0.9249&	 1.0938&	 ${-0.8475}$&	 $0.7008\times 10^{-3}$&	 $0.9802\times 10^{-3}$\\
100&	 0.025&	 0.9058&	 1.1275&	 ${-0.9197}$&	 $0.9410\times 10^{-3}$&	 $1.4579\times 10^{-3}$\\
100&	 0.03&	 0.8875&	 1.1658&	 ${-1.0056}$&	 $1.0896\times 10^{-3}$&	 $1.8800\times 10^{-3}$\\
100&	 0.035&	 0.8703&	 1.2088&	 ${-1.1066}$&	 $1.0770\times 10^{-3}$&	 $2.0778\times 10^{-3}$\\
100&	 0.04&	 0.8545&	 1.2567&	 ${-1.2245}$&	 $0.8363\times 10^{-3}$&	 $1.8089\times 10^{-3}$\\
100&	 0.045&	 0.8403&	 1.3094&	 ${-1.3609}$&	 $0.3070\times 10^{-3}$&	 $0.7455\times 10^{-3}$\\
100&	0.04701&	0.8350&	1.3320&		${-1.4215}$&	0&	0\\ 
\hline
400&	 0.01&	 0.9508&	 1.0574&	 ${-0.7783}$&	 $0.2949\times 10^{-3}$&	 $0.3648\times 10^{-3}$\\
400&	 0.015&	 0.9177&	 1.1091&	 ${-0.8894}$&	 $0.5922\times 10^{-3}$&	 $0.8648\times 10^{-3}$\\
400&	 0.02&	 0.8847&	 1.1771&	 ${-1.0451}$&	 $0.7686\times 10^{-3}$&	 $1.3606\times 10^{-3}$\\
400&	 0.025&	 0.8556&	 1.2599&	 ${-1.2485}$&	 $0.6481\times 10^{-3}$&	 $1.4052\times 10^{-3}$\\
400&	 0.03&	 0.8324&	 1.3559&	 ${-1.5026}$&	 $0.1290\times 10^{-3}$&	 $0.3422\times 10^{-3}$\\
400&	0.03083&	0.8292&	1.3730&		${-1.5500}$&	0&	0
\end{tabular}
\caption{Computed values for the fundamental mode.}
\label{table freq node=0}
\end{table*}

\setlength{\tabcolsep}{8pt}
\begin{table*}
\begin{tabular}{lllllll}
$\eta/(\mu^2/m_P^2)$ & $\phi_0(0)/m_P$ &$\omega/\mu$ & $\hat{\omega} / \mu$ & $\zeta_1/\mu^2$ &  $\chi^2/\mu^2$ & $\hat{\chi}^2/\mu^2$ \\
\hline
\hline
0&	 0.04&	 0.8840&	 1.1645&	 ${-1.0042}$&	 $0.7892\times 10^{-2}$&	 0.0137\\
0&	 0.06&	 0.8415&	 1.2806&	 ${-1.2935}$&	 $1.5174\times 10^{-2}$&	 0.0351\\
0&	 0.08&	 0.8085&	 1.4299&	 ${-1.7291}$&	 $2.2766\times 10^{-2}$&	 0.0712\\
0&	 0.1&	 0.7851&	 1.6245&	 ${-2.4085}$&	 $2.9433\times 10^{-2}$&	 0.1260\\
0&	 0.12&	 0.7713&	 1.8824&	 ${-3.5133}$&	 $3.4057\times 10^{-2}$&	 0.2028\\
0&	 0.14&	 0.7677&	 2.2305&	 ${-5.3947}$&	 $3.5702\times 10^{-2}$&	 0.3014\\
0&	 0.16&	 0.7743&	 2.7081&	 ${-8.7602}$&	 $3.3792\times 10^{-2}$&	 0.4134\\
0&	 0.18&	 0.7904&	 3.3722&	 ${-15.0690}$&	 $2.8409\times 10^{-2}$&	 0.5171\\
0&	 0.2&	 0.8130&	 4.3002&	 ${-27.3010}$&	 $2.0283\times 10^{-2}$&	 0.5675\\
0&	 0.22&	 0.8357&	 5.5908&	 ${-51.3178}$&	 $0.8862\times 10^{-2}$&	 0.3966\\
0&	0.2301&	0.8447&	6.4205&	${-71.2434}$&	0&	0\\
\hline
100&	 0.04&	 0.8545&	 1.2567&	 ${-1.2462}$&	 $1.0038\times 10^{-2}$&	 0.0217\\
100&	 0.06&	 0.8081&	 1.4973&	 ${-1.9525}$&	 $1.7335\times 10^{-2}$&	 0.0595\\
100&	 0.08&	 0.7883&	 1.8198&	 ${-3.1402}$&	 $2.1562\times 10^{-2}$&	 0.1149\\
100&	 0.1&	 0.7884&	 2.2336&	 ${-5.0970}$&	 $2.1982\times 10^{-2}$&	 0.1764\\
100&	 0.12&	 0.8017&	 2.7581&	 ${-8.3403}$&	 $1.9234\times 10^{-2}$&	 0.2277\\
100&	 0.14&	 0.8218&	 3.4251&	 ${-13.8187}$&	 $1.4467\times 10^{-2}$&	 0.2513\\
100&	 0.16&	 0.8431&	 4.2819&	 ${-23.2943}$&	 $0.8472\times 10^{-2}$&	 0.2185\\
100&	 0.18&	 0.8601&	 5.3982&	 ${-40.1006}$&	 $0.0267\times 10^{-2}$&	 0.0105\\
100&	0.1805&	0.8604&	5.4310&	${-40.6736}$&	0&	0\\
\hline
400&	 0.04&	 0.8039&	 1.5810&	 ${-2.2058}$&	 $1.0542\times 10^{-2}$&	 0.0408\\
400&	 0.06&	 0.7964&	 2.1377&	 ${-4.4044}$&	 $1.1507\times 10^{-2}$&	 0.0829\\
400&	 0.08&	 0.8185&	 2.8204&	 ${-8.1458}$&	 $0.8650\times 10^{-2}$&	 0.1027\\
400&	 0.1&	 0.8453&	 3.6412&	 ${-14.3217}$&	 $0.4448\times 10^{-2}$&	 0.0825\\
400&	0.1186&	0.8632&	4.5597&	${-23.6175}$&	0&	0
\end{tabular}
\caption{Computed values for the first excited mode.}
\label{table freq node=1}
\end{table*}

\section{Comparison with literature}
\label{sec:comparsion}

In this appendix, we compare some of our results from Sec.\ \ref{sec:radial} with those found in the literature.  In particular, Gleiser and Watkins \cite{Gleiser:1988ih} and Hawley and Choptuik \cite{Hawley:2000dt} have reported results for $\eta = 0$, i.e.\ in the absence of self-interactions.

We have computed the critical value $\phi_0(0)/m_P = 0.2301$ for the first excited mode with $\eta = 0$ (see Table \ref{table freq node=1}).  This value is computed in both \cite{Gleiser:1988ih, Hawley:2000dt}.  These two papers report, respectively, 1.16 and (approximately) 1.15, but they scale $\phi_0(0)/m_P$ with a factor of $\sqrt{8\pi}$, and thus $1.16/\sqrt{8\pi} = 0.2314$ and $1.15/\sqrt{8\pi} = 0.2294$, which is consistent with our result.

\setlength{\tabcolsep}{8pt}
\begin{table*}
\begin{tabular}{llllll||ll}
$\phi_0(0)/(m_P/\sqrt{8\pi})$ & $\phi_0(0)/m_P$ & $\omega/\mu$ & $\hat{\omega} / \mu$ &  $\chi^2/\mu^2$ & $\hat{\chi}^2/\mu^2$ & \cite{Hawley:2000dt}:\ $\hat{\omega}/\mu$ & \cite{Hawley:2000dt}:\ $\hat{\chi}^2/\mu^2$ \\
\hline
\hline
0.06&	0.01197&	0.9606&	1.0417&	$0.2307\times10^{-3}$&	$0.2713\times 10^{-3}$&  1.0417& $0.28\times10^{-3}$ \\
0.1&	0.01995&	0.9367&	1.0727&		$0.5121\times10^{-3}$&	$0.6717\times 10^{-3}$&  1.0727& $0.67\times10^{-3}$ \\
0.14&	0.02793&	0.9144&	1.1067&		$0.7594\times10^{-3}$&	$1.1112\times 10^{-3}$& 1.1067& $1.11\times10^{-3}$ \\
0.18&	0.03590&	0.8939&	1.1439&		$0.8628\times10^{-3}$&	$1.4132\times 10^{-3}$& 1.1440& $1.41\times10^{-3}$ \\
0.22&	0.04388&	0.8749&	1.1849&		$0.7162\times10^{-3}$&	$1.3135\times 10^{-3}$& 1.1849& $1.31\times10^{-3}$ \\
0.26&	0.05186&	0.8576&	1.2299&		$0.2151\times10^{-3}$&	$0.4424\times 10^{-3}$&  1.2299& $0.45\times10^{-3}$ \\
0.27&	0.05386&	0.8535&	1.2419&		$0.2217\times10^{-4}$&	$0.0469\times 10^{-3}$&  1.2419& $0.05\times10^{-3}$ 
\end{tabular}
\caption{Comparison of values we have computed with those computed in \cite{Hawley:2000dt} for the fundamental mode.}
\label{table fundamental}
\end{table*}

\begin{table*} 
\begin{tabular}{llllll||ll}
$\phi_0(0)/(m_P/\sqrt{8\pi})$ & $\phi_0(0)/m_P$ & $\omega/\mu$ & $\hat{\omega} / \mu$ & $\chi^2/\mu^2$ & $\hat{\chi}^2/\mu^2$ & \cite{Hawley:2000dt}:\ $\hat{\omega}/\mu$ & \cite{Hawley:2000dt}:\ $\hat{\chi}^2/\mu^2$ \\
\hline
\hline
0.6&	0.1197&	0.7715&	1.8777&	0.03401&	0.2015& 1.8777 & 0.22 \\
0.7&	0.1396&	0.7676&	2.2230&		0.03570&	0.2994& 2.2230 & 0.32 \\
0.8&	0.1596&	0.7740&	2.6963&		0.03387&	0.4110& 2.6963 & 0.43 \\
0.9&	0.1795&	0.7899&	3.3536&		0.02857&	0.5150& 3.3536 & 0.53 \\
1&		0.1995&	0.8123&	4.2714&		0.02053&	0.5676& 4.2714 & 0.54 \\
1.1&	0.2194&	0.8351&	5.5470&		0.009283&	0.4096& 5.5471 & 0.42 \\
1.12&	0.2234&	0.8390&	5.8554&		0.006222&	0.3031& 5.8555 & 0.305 \\
1.14&	0.2274&	0.8425&	6.1841&		0.002705&	0.1457& 6.1842 & 0.146 \\
1.15&	0.2294&	0.8442&	6.3566&		0.000746&	0.04233& 6.3566 & 0.0430 
\end{tabular}
\caption{Comparison of values we have computed with those computed in \cite{Hawley:2000dt} for the first excited mode.}
\label{table excited}
\end{table*}

Hawley and Choptuik list a number of oscillation frequencies for the fundamental and first excited modes in Appendix A of \cite{Hawley:2000dt}.  It is important to note that they list what we have labeled as $\hat{\omega}/ \mu$ and $\hat{\chi}^2/\mu^2$ in the main text.  In Table \ref{table fundamental}, we list the results reported in \cite{Hawley:2000dt} for the fundamental mode and the corresponding values we have computed.  As can be seen, there is excellent agreement between $\hat{\omega}/\mu$ and $\hat{\chi}^2/\mu^2$.

In Table \ref{table excited} we do the same but for the first excited mode.  We find again excellent agreement for $\hat{\omega}/\mu$.  However, we find some discrepancy with $\hat{\chi}^2/\mu^2$, although our numbers are mostly within the reported error for theirs (which is $\pm 2$ in the final digit).  Although our numbers are mostly consistent with theirs, there are two reasons why we believe our numbers to be more accurate.  The first is that we compute both $\hat{\omega}$ and $\zeta_1$ to near machine-precision accuracy, with $\zeta_1$ computed to this accuracy for each value of $\hat{\chi}^2$ that is tried in the shooting method.  $\hat{\chi}^2$ is then computed to an accuracy greater than listed in the tables.  The second reason is that we have confirmed the accuracy of our numerical method for computing frequencies by comparing values to dynamical solutions, as explained in Sec.\ \ref{sec:dynamic}.


\section{Units}
\label{app:units}

Figures \ref{fig:static_MRomega}--\ref{fig:dynamic} and Tables \ref{table freq node=0}--\ref{table excited} presented results in terms of dimensionless quantities.  The benefit in doing this is that the results are valid for an arbitrary scalar field mass, $\mu$, and $\mu$ does not have to be specified.  In this appendix, we list a few astrophysical\hyp friendly unit conversions, which do require specification of $\mu$.

The boson star mass was given in terms of $M / (m_P^2/\mu)$.  To convert this to solar masses (M$_\odot$),
\begin{equation}
M = 1.3360 \left( \frac{10^{-10}\text{ eV}}{\mu} \right) \left[ M / (m_P^2/\mu) \right] \quad \text{[M$_\odot$]} .
\end{equation}
The boson star radius was given in terms of $\mu R_{95}$.  To convert this to kilometers (km),
\begin{equation}
R_{95} =
1.9732  \left( \frac{10^{-10}\text{ eV}}{\mu} \right) (\mu R_{95}) \quad \text{[km]} .
\end{equation}
The oscillation frequency was given in terms of $\omega/\mu$.  To convert this to kilohertz (kHz),
\begin{equation}
\omega = 151.93 \left(\frac{\mu}{10^{-10}\text{ eV}} \right) (\omega/\mu) \quad \text{[kHz]}.
\end{equation}
Similarly, the squared radial oscillation frequency was given in terms of $\chi^2/\mu^2$.  To convert this to squared kilohertz (kHz$^2$),
\begin{equation}
\chi^2 = 23083 \left(\frac{\mu}{10^{-10}\text{ eV}} \right)^2 (\chi^2/\mu^2) \quad \text{[kHz$^2$]}.
\end{equation}





\end{document}